\documentclass[aps,prb,floatfix,showpacs,superscriptaddress]{revtex4}
\usepackage{epsfig}
\usepackage{bm}
\usepackage{latexsym}

\begin{document}

\newcommand{\td}{\tilde{\tau}_w}
\newcommand{\Pt}{{\mathcal P}_{w}(\tilde{\tau}_w)}
\newcommand{\ptime}{\tilde{\tau}}
\newcommand{\Ppt}{{\cal P}_{\tilde{\tau}}(\tilde{\tau})}

\title{Statistics of delay times in mesoscopic systems
as a manifestation of eigenfunction fluctuations}
\author{Alexander Ossipov}
\email{aossipov@ictp.trieste.it}
\affiliation{The Abdus Salam International Centre for Theoretical Physics,
Strada Costiera 11, 34014 Trieste, Italy}
\author{Yan V. Fyodorov}
\email{yan.fyodorov@brunel.ac.uk} \affiliation{Department of
Mathematical Sciences, Brunel University, Uxbridge UB83PH, United
Kingdom}

\date{\today}

\begin{abstract}
We reveal a general explicit relation between the statistics of
delay times in one-channel reflection from a mesoscopic sample of
any spatial dimension and the statistics of the eigenfunction
intensities in its closed counterpart. This opens a possibility to
use experimentally measurable delay times as a sensitive probe of
eigenfunction fluctuations. For the particular case of quasi-one
dimensional geometry of the sample we use an alternative technique
to derive the probability density of partial delay times for any
number of open channels.
\end{abstract}
\pacs{74.78.Na, 05.45.Mt, 73.23.-b, 42.25.Bs
}

\maketitle
\section{Introduction}

The standard way of describing scattering in quantum systems
employs a $M\times M$ unitary matrix known as $S$-matrix, with $M$
standing for the number of scattering channels open at a given
energy. Then the Wigner delay time \cite{W55} defined as the
energy derivative of the total scattering phaseshift:
$\tau_w=-i\hbar \frac{\partial}{\partial E} {\rm ln Det S}$ is one
of the most important and frequently used characteristics of
quantum scattering, see e.g.\cite{CN,Ianna}. It can be interpreted
as a time delay in propagation of the peak of the wave packet due
to scattering in comparison with a free wavepacket propagation.
More detailed
 characterization of scattering calls for introducing the
 Wigner-Smith time delay matrix\cite{S60}
$Q=i\hbar\frac{\partial S^\dagger}{\partial E}S$, whose
eigenvalues are frequently called {\it proper} delay times.
Alternatively, denoting $M$ unimodular eigenvalues of the
$S-$matrix as $\{\exp(i\theta_1),\dots,\exp(i\theta_M)\}$, one can
consider {\it partial} delay times defined\cite{Fyodorov1997} as
$\tau_a=\partial \theta_a /\partial E, \; a=1,\dots, M$. Beyond
one-channel case proper and partial delay times differ, although
the sum of partial/proper delay times over all $M$ scattering
channels are always equal and yield the Wigner delay time.

Statistics of delay times of all sorts were studied intensively in
the framework of quantum chaotic scattering. Earlier works on the
subject used various approximation schemes, frequently not very
well controlled\cite{early}. It is however well known that
 quantum statistical properties of classically chaotic systems are to large
extent universal and independent of their microscopic nature
\cite{Haake,St99} which allows one to use the Random Matrix Theory
(RMT) for description of their properties\cite{Guhr}. When applied
to the scattering problems and in particular to the study of delay
times statistics the RMT yielded many exact analytical results in
recent years,
see\cite{Fyodorov1997,L95,Fyodorov1997b,Gopar1996,B97,Gopar1998,FSS97,
SFS001,FO04,SS2003} and references therein. Those were
successfully verified in numerical simulations of several quantum
chaotic systems of quite a diverse nature, see e.g. \cite{tdnum}.
Some properties of delay times were also studied by semiclassical
methods\cite{semi}.

In the general framework of mesoscopic systems the RMT regime
corresponds to the so called zero-dimensional limit. In this limit
the closed counterpart of a scattering system is characterized by
the only non-trivial energy scale which is simply the mean level
spacing $\Delta$, and RMT predictions for statistics of
eigenfunctions and energy levels were successfully tested
experimentally in microwave and acoustic experiments\cite{exp0D}.

Beyond the zero-dimensional limit the particular nature of the
sample geometry and type of disorder causing chaotic behaviour
starts to play an important role. In particular, in systems with
random short-range impurities another energy scale known as the
Thouless energy $E_c$ plays a prominent role, the ratio $g=2\pi
E_c/\Delta$ being known as the dimensionless conductance of the
sample. This parameter controls the system properties, with the
universal RMT regime being recovered in the limit $g\to \infty$.
 The statistical properties of the energy
spectra and eigenfunctions of closed mesoscopic systems as a
function of dimensional conductance were under intensive
investigation for more than a decade, both analytically
\cite{anal,FM94,multif,bla,prig}, numerically\cite{num,MEM}, and
some predicted features are also seen in microwave
experiments\cite{expand}. The research resulted in a detailed
picture emerging for various regimes, including the most difficult
case of the Anderson localization transition, see the review
\cite{M00} for a general picture. At the same time systematic
statistical analysis of delay times beyond the universal
zero-dimensional limit is still lacking, apart from a few useful
insights in \cite{JVK89,Schomerus,SOKG00,KW02,OKG03,F03}.

The goal of the present paper is to initiate systematic
investigation of delay times beyond the zero-dimensional
approximation in metallic, critical and localized regimes. To this
end, in the first part of the paper we reveal a very general
relation between the delay time distribution and the distribution
of eigenfunction intensities for a single channel scattering. On
one hand, this relation allows us to use the existent knowledge on
eigenfunction statistics to provide explicit expressions for delay
times distributions in various regimes of interest. On the other
hand, since phase shifts and delay times are experimentally
measurable quantity, especially in the one-channel reflection
experiment \cite{Doron,GSST82,JPM03,Kuhl2004}, the relation opens
a new possibility for experimental study of eigenfunctions.

In the second part of the paper we consider a model of a
quasi-one-dimensional disordered wire represented by Banded Random
Matrix (BRM) ensemble. It is well-known that in the limit of the
thick wire this model is exactly solvable \cite{FM94Int} by the
transfer matrix approach \cite{EL83}. The same technique when
combined with methods introduced in\cite{Fyodorov1997} for the
scattering problem yields the distribution of partial delay times
for any finite number of open channels. In particular, for the
single open channel it reproduces the result obtained from the
general relation developed in the first section.

\section{Delay time distribution for one-channel scattering:
relation to eigenfunction statistics.}

Consider a single-channel antenna/lead perfectly coupled to an
absorptive disordered system, characterized by the (energy,
absorption and position dependent) reflection coefficient
$R=R(\eta,{\bf r})\equiv |S(E,\eta)|^2\le 1$, where ${\bf r}$
stands for the coordinate of antenna port, and $\eta\ge 0$ for the
absorption parameter. Here and henceforth the absorption is
considered as uniform, position-independent process and therefore
can be accounted for as an imaginary shift of the energy of
incoming waves. For the definition of "perfect coupling" see the
equation Eq.~(\ref{coupling}) below and the discussion preceding it.

The Wigner delay time can be conveniently written as
$\tau_w=\lim_{\eta\to 0}\frac{1-R}{2\eta}$, see e.g.\cite{Doron}.
Introduce variable $x\ge 1$ via $R=\frac{x-1}{x+1}$, and let
${\cal P}^{(\eta)}_0 (x)$ stands for the distribution function of
$x$ at given absorption.

In the limit of zero absorption $\eta\to 0$ the reflection
coefficient $R\to 1$, hence $x\to \infty$. In fact, correct limit
is such that $x=\frac{z}{\eta},\,\, z<\infty$. The variable $z$ is
nothing else but the inverse delay time:
\begin{equation}\label{1}
\frac{1}{\tau_w}=\lim_{\eta\to 0}\frac{2\eta}{1-R}=\lim_{\eta\to
0}\eta (x+1)=z
\end{equation}
The distribution function function ${\cal P}^{(\eta)}_0 (x)$
behaves in the limit of vanishing absorption as $\tilde{{\cal
P}}_0 (z/\eta)\to \eta \tilde{{\cal P}}_0 (z)$, where
$\tilde{{\cal P}}_0 (z)$ is a well-defined distribution function
of variable $z$ (i.e. of the inverse delay time).

From the other hand,  let $ G(E+i\eta; {\bf r}; {\bf r}')$ be the
Green's function of a closed system with broadening $\eta$, so
that the local density of states (LDoS) at any point ${\bf r}$ in
the sample is given by
\begin{equation}
\tilde{v}_{\eta}({\bf r})=-\frac{1}{\pi}\mbox{Im}\, G(E+i\eta;{\bf
r}; {\bf r})=\frac{1}{\pi} \sum_{n=1}^N |\psi_n({\bf
r})|^2\frac{\eta}{(E-E_n)^2+\eta^2}
\end{equation}
Here $\psi_n({\bf r})$ stands for the local amplitude of the
eigenfunction of the wave/Schr\"{o}dinger operator describing a
classical wave/quantum particle in the disordered sample and
corresponding to the eigenfrequency/energy level $E_n$.

Using this relation it is easy to satisfy oneself that for any
integer $k=1,2,3,\ldots$ holds
\begin{equation}\label{2}
\lim_{\eta\to 0}\,[\eta^{k-1} \pi^k\tilde{v}^{k}_{\eta}({\bf r})]=
\frac{\sqrt{\pi} \Gamma(k-\frac{1}{2})}{\Gamma(k)}\sum_{n=1}^N
|\psi_n({\bf r})|^{2k}\delta(E-E_n)
\end{equation}
where $\Gamma(z)$ stands for the Euler gamma-function.

 Indeed, in the limit $\eta\to 0$ the
nonvanishing contribution to the above expression comes only from
the most singular ("diagonal") term in the product of sums over
eigenvalues:
$$
 \eta^{k-1}\sum_{n=1}^N |\psi_n({\bf r})|^{2k}\frac{\eta^{k}}{[(E-E_n)^2+\eta^2]^k}
$$
Evaluating the standard integrals, we observe that:
\[
\int_{-\infty}^{\infty}dx \frac{
\eta^{2k-1}}{[x^2+\eta^2]^k}=\int_{-\infty}^{\infty}du
\frac{1}{[u^2+1]^k}=\frac{\sqrt{\pi}\Gamma(k-\frac{1}{2})}{\Gamma(k)}
\]
which yields the above result Eq.(\ref{2}).

The right hand side of Eq.(\ref{2}) is proportional to the moments
of the local eigenfunction intensity $y=V|\psi_n({\bf r})|^{2}$ ,
with $V$ standing for the volume of our sample, averaged in a
small energy window around point $E$ of the spectrum. Denoting
additional disorder averaging by brackets and introducing the
distribution function ${\cal P}_v (v)$ of the random variable
$v_{\eta}({\bf r})= \tilde{v}_{\eta}({\bf r})V\Delta$ , with
$\Delta$ being the mean level spacing of the sample,
 we therefore can rewrite our relation as:
\begin{equation}\label{3}
\langle  y^k \rangle=\frac{\sqrt{\pi}\Gamma(k)}{\Gamma(k-\frac{1}{2})}
\lim_{\eta\to 0}\,\left(\frac{\eta\pi}{\Delta}\right)^{k-1}\int_0^{\infty }
dv \, v^k\, {\cal P}_v (v)
\end{equation}

At the next step we use a recently discovered relation between the
probability distribution ${\cal P}_v (v)$ of LDoS and the function
${\cal P}^{(\eta)}_0(x),\,\, x\ge 1$ \cite{FyodorovSavin04}:
\begin{equation}\label{P(v)}
{\mathcal P}_{v}(v) = \frac{\sqrt{2}}{\pi
v^{3/2}}\int_0^{\infty}\!\!dq\, {\cal
P}^{(\eta)}_0\left[q^2+\frac{1}{2}\Bigl(v+\frac{1}{v}\Bigr)\right]\,,
\end{equation}
valid for those disordered and chaotic systems which allow an
effective description in terms of the nonlinear $\sigma-$ model.
The existence of this relation is based on the two following
observations. First, the scattering matrix for a one-channel
scattering can be expressed in terms of the diagonal matrix
element of the resolvent of the corresponding closed system (see
Eqs.(\ref{S-K},\ref{K-matr})
 in the next section). In this way one can relate the
 probability distribution of LDoS and statistical properties
 of the scattering matrix. The
second observation is that under conditions of perfect coupling
the phase of the $S$-matrix is statistical independent of the
$S$-matrix modulus $R=\frac{x-1}{x+1}$ and is distributed
uniformly over the unit circle. Consequently, all non-trivial
information on the scattering matrix is contained in the
distribution function of its modulus $R$, or equivalently in the
distribution function of $x$. This fact explains the appearance of
the function ${\cal P}^{(\eta)}_0(x)$  on the right-hand side of
Eq.(\ref{P(v)}) , see \cite{FyodorovSavin04} for the detailed
derivation and discussion. It is important also to remember that
the above distribution ${\mathcal P}_{v}(v)$ is normalized in such
a way that the first moment is equal to unity: $\langle v\rangle
\equiv 1$.

Rescaling  the variable  $v=z/\eta$ in the integral (\ref{3}) we
see that it is $\lim_{\eta\to 0} {\mathcal P}_{v}(z/\eta)/ \eta^2$
which we are interested in. The limit can be calculated from
(\ref{P(v)}) by changing the variable in the integral to
$q=t/\sqrt{\eta}$:
\begin{equation}
\lim_{\eta\to 0} \frac{1}{\eta^2}{\mathcal P}_{v}\left(\frac{z}{\eta}\right)=
 \frac{\sqrt{2}}{\pi z^{3/2}}\int_0^{\infty}\!\!dt\;
\tilde{{\cal P}}_0 \left(t^2+\frac{z}{2}\right)
\end{equation}
In this way we bring Eq.(\ref{3}) to the form
\begin{equation}
\label{y^k}
\langle  y^k \rangle=\left(\frac{\pi}{\Delta}\right)^{k-1}
\sqrt{\frac{2}{\pi}}\frac{\Gamma(k)}
{\Gamma(k-\frac{1}{2})}
\int_0^{\infty}\!\!dz\;z^{k-3/2}\int_0^{\infty}\!\!dt\; \tilde{{\cal P}}_0
\left(t^2+\frac{z}{2}\right)
\end{equation}
Introducing the new variable $s=t^2+z/2$ we further notice that
\begin{eqnarray}
\int_0^{\infty}\!\!dz\;z^{k-3/2}\int_0^{\infty}\!\!dt\; \tilde{{\cal P}}_0
\left(t^2+\frac{z}{2}\right) = \int_0^{\infty}\!\!ds\;\tilde{{\cal P}}_0 (s)
\int_0^{2s}\!\!dz\; \frac{z^{k-3/2}}{2\sqrt{s-\frac{1}{2}z}}=\nonumber\\
=\sqrt{\frac{\pi}{2}}\frac{\Gamma(k-\frac{1}{2})}{\Gamma(k)}\int_0^{\infty}\!\!ds\;\tilde{{\cal
P}}_0 (s) (2s)^{k-1}
\end{eqnarray}
Substituting now the last expression into (\ref{y^k}) we arrive at
the relation:
\begin{equation}\label{8}
\int_0^{\infty } dy \, {\cal P}_y (y)\, y^{k}\equiv \langle y^k
\rangle =\int_0^{\infty } ds \, \tilde{\cal P}_0 (s)\, \left(\frac{2\pi s}
{ \Delta}\right)^{k-1}
\end{equation}
valid for all integer $k\ge 1$. We therefore conclude that the
distribution functions in the left and right hand sides of this
expression are simply related to each other by ${\cal P}_y
(\tilde{z})=\tilde{\cal P}_0 (\tilde{z})\, \tilde{z}^{-1}$, where
$\tilde{z}=2\pi z/\Delta$.
 Remembering interpretation of $z$ as the
 inverse delay time and introducing the distribution
$\tilde{\cal P}_{w} (\tilde{\tau}_w)$ of scaled time delays
$\tilde{\tau}_w=\tau\Delta/2\pi$ we finally arrive at the
following simple but fundamental relation between moments of
eigenfunction intensity $y$ and the inverse moments of the time
delay:
\begin{equation}\label{9a}
\left\langle \tilde{\tau}_w^{-k}\right\rangle=\left\langle
y^{k+1}\right\rangle
\end{equation}
resulting in the functional relation between the two
distributions:
\begin{equation}\label{9}
\tilde{\cal P}_{w}
(\tilde{\tau}_w)=\frac{1}{\tilde{\tau}_w^3}{\cal P}_y
\left(\frac{1}{\tilde{\tau}_w}\right),
\end{equation}
and constituting one of the central results of the present paper.
Some remarks on the conditions of validity of Eqs.(\ref{9a},
\ref{9}) are given in the last section of the present paper.

Let us now demonstrate how the relation Eq.(\ref{9}) works in
various situations. Let us start with the case of a
``zero-dimensional'' system which can be described by the RMT. The
distributions of eigenfunction intensities typical for the
zero-dimensional case and various symmetry classes characterized
by $\beta=1,2,4$ \cite{Guhr} are given by the so-called
$\chi-$squared distribution:
\begin{equation}\label{chi2}
{\cal P}_y(y)=C_{\beta} y^{\beta/2-1}e^{-\frac{\beta}{2}y},\quad
C_{\beta}=(\beta/2)^{\beta/2}/\Gamma(\beta/2)
\end{equation}
 with particular case $\beta=1$ frequently being referred to as
the Porter-Thomas distribution. The known expressions
\cite{Fyodorov1997b,Gopar1996} for all three pure symmetry classes
immediately follow as:
\begin{equation}\label{rmt}
{\mathcal P}_{w}(\tilde{\tau}_w)=[(\beta/2)^{\beta/2}/\Gamma(\beta/2)]
\tilde{\tau}_w^{-\beta/2-2}e^{-\beta/2\tilde{\tau}_w}
\end{equation}
 Moreover, the distribution of
the one-channel delay times in the crossover regime between
unitary ($\beta=2$) and orthogonal ($\beta=1$) symmetry classes
was calculated in Ref.\cite{FSS97} :
\begin{eqnarray}\label{cross}
{\mathcal P}_{w}(\tilde{\tau}_w)&=&\frac{1}{2\td^3}\int_{-1}^{1}d\lambda\;
\int_{1}^{\infty}d\lambda_2\:\lambda_2^2 e^{-X^2(\lambda_2^2-1)}e^{-
\lambda_2^2/\td}I_0\left[\frac{\lambda_2\sqrt{\lambda_2^2-1}}{\td}\right]
{\cal T}_2(\lambda,\lambda_2),\\
{\cal T}_2(\lambda,\lambda_2)&=&
2X^2\left[(1-\lambda^2)e^{-\alpha}+\lambda_2^2
(1-e^{-\alpha})\right]-(1-e^{-\alpha}),\nonumber\\
\end{eqnarray}
where $\alpha=X^2(1-\lambda^2)$, $I_0(z)$ stands for the modified
Bessel function, and $X$ is a crossover driving parameter
\cite{footnote}. This result can be easily recovered from the
distribution of the eigenfunction intensities in the crossover
found in Ref.\cite{FE94}:
\begin{eqnarray}
{\cal P}_y(y)&=& 2 \int_1^{\infty}dt\:\left\{\Phi_1(X)+\left[(tX)^2-1\right]
\Phi_2(X)\right\}(tX)^2 I_0(yt\sqrt{t^2-1})e^{-yt^2-X^2(t^2-1)},\\
&&\Phi_1(X)=\frac{e^{-X^2}}{X}\int_0^Xdx\:e^{x^2}, \hspace{1.2cm}
\Phi_2(X)=\frac{1-\Phi_1(X)}{X^2}.\nonumber
\end{eqnarray}
In the metallic regime beyond RMT the perturbative corrections to
the body of the distribution of eigenfunction intensities were
calculated using supersymmetric non-linear $\sigma$-model in Ref.
\cite{FM94}. Then relation (\ref{9}) yields the following
distributions of the Wigner delay times:
\begin{eqnarray}
\label{metallic}
{\mathcal P}_{w}(\tilde{\tau}_w)&=&\frac{e^{-1/2\td}}{\sqrt{2\pi}\td^{5/2}}
\left[1+\frac{\kappa}{2}\left(\frac{3}{2}-\frac{3}{\td}+\frac{1}{2\td^2}\right)+
\dots\right]\;\;\; \beta=1,\nonumber\\
\Pt &=&\frac{e^{-1/\td}}{\td^3}
\left[1+\frac{\kappa}{2}\left({2}-\frac{4}{\td}+\frac{1}{\td^2}\right)+\dots
\right]\;\;\; \beta=2,\nonumber\\
\Pt &=&\frac{4e^{-2/\td}}{\td^4}
\left[1+\frac{\kappa}{2}\left({3}-\frac{6}{\td}+\frac{2}{\td^2}\right)+\dots
\right]\;\;\; \beta=4,
\end{eqnarray}
Here  the parameter $\kappa=a/g$ is just the diagonal part of the
so-called diffusion propagator $\Pi({\bf r},{\bf
r})$\cite{FM94,M00} and is inversely proportional to the
dimensionless conductance $g$. The exact value of the constant $a$
depends essentially on the sample geometry and on the coordinates
of the lead.

Eq.(\ref{metallic}) holds for relatively large delay times $\td \gtrsim
 \sqrt{\kappa}$, while in the opposite case  the distribution is dominated by
the existence of the anomalously localized states (see \cite{M00} for a
 review) and has the following  behavior for dimensionality $d=2,3$
\cite{multif}:
\begin{eqnarray}
\Pt &\sim& \exp\left(\frac{\beta}{2}\left\{-\frac{1}{\td}+\kappa
\frac{1}{\td^2}+\dots\right\}\right), \;\;\;\; \kappa\lesssim \td
\lesssim \sqrt{\kappa},\\
\Pt &\sim& \exp({-C_d \ln^d(1/\td)}), \;\;\;\; \td \lesssim \kappa.
\end{eqnarray}
It deserves mentioning that the log-normal distribution found in
Ref. \cite{OKG03} was observed for the opposite limit of {\it
large} delay times and {\it  many} open channels.

Another important consequence of (\ref{9}) is that the
eigenfunction multifractality typical for the vicinity of the
Anderson localization transition \cite{M00} reflects itself in the
distribution of Wigner delay times. Generally this means that the
negative moments of the delay time scale anomalously with the
system size $L$:
\begin{equation}
\label{scale}
\left<\frac{1}{\tau_w^{q-1}}\right>  \sim L^{-D_q (q-1)}
\end{equation}
where $D_q$ is a fractal dimension of the eigenfunctions. The idea
of the anomalous scaling (\ref{scale}) of the time delays was
suggested by one of the authors in Ref.\cite{F03}.

The fractal dimensions of the eigenfunctions is known analytically
for a  two-dimensional system in the metallic regime, where the
deviation from the normal scaling is determined by the inverse
conductance (the regime of weak multifractality) \cite{M00}:
\begin{equation}
\label{weak}
\left<\frac{1}{\tau_w^{q-1}}\right> \sim L^{-(2-\frac{q}{\beta\pi g}) (q-1)},
\;\;\;\;\; q\ll 2\beta \pi g.
\end{equation}
another case when the anomalous dimensions are known analytically
is the model of power-law random banded matrices, whose elements
are independent random variables $H_{ij}$ with the variance
decreasing in a power-law fashion: $\left<(H_{ij})^2\right>=
[1+(|i-j|/b)^{2\alpha}]^{-1}$. For $\alpha=1$ this model shows
critical
 behavior and the fractal dimensions of the eigenfunctions can be calculated
for $b\gg 1$ \cite{MFD96}:
\begin{equation}
\label{pl}
\left<\frac{1}{\tau_w^{q-1}}\right> \sim L^{-(1-\frac{q}{2\beta\pi b}) (q-1)}.
\end{equation}

\section{Partial delay times in a quasi-one-dimensional geometry}

In this section we consider a quasi-one-dimensional disordered
sample of length $L$ with a perfect lead attached to one of its
edges, the other edge being impenetrable for the waves. The lead
supports $M$ scattering channels, thus the $S$-matrix is $M\times
M$ unitary matrix  and the partial delay times are defined in the
standard way as $\tau_a=\partial \theta_a /\partial E, \;
a=1,\dots, M$.

Following the method suggested for the zero-dimensional case in
\cite{Fyodorov1997b,Fyodorov1997} we address the statistics of the
partial delay times by considering the two-point correlation
function of eigenvalue densities of the $K$-matrix. The latter
matrix is Hermitian and is defined in terms of the scattering
matrix via the relation
\begin{equation}
\label{S-K}
\hat{S}=\frac{\hat{I}-i\pi\hat{K}}{\hat{I}+i\pi\hat{K}}.
\end{equation}

Denoting the real eigenvalues of the $K$-matrix by
$z_a,\;a=1,\dots,M$ the partial delay times can be obviously
written as $\tau_a(E)=\partial \theta_a (E)
 /\partial E=-\frac{2}{1+z_a^2(E)}\frac{\partial z_a (E)}{\partial E}$. Knowing the
joint probability density of random variables $z_a$ and
$v_a=\partial z_a /\partial E$
\begin{equation}
{\cal P}_E(z,v)=\frac{1}{M}\left<\sum_{a=1}^{M}\delta(z-z_a)\;\delta\left(v-
\frac{\partial z_a}{ \partial E}\right)\right>
\end{equation}
one can recover the mean density of the partial delay times as
\begin{equation}
{\cal P}_{\tau}(\tau)=\frac{1}{M}\left<\sum_{a=1}^{M}\delta(\tau-\tau_a)\right>
=\int_{-\infty}^{\infty}\int_{-\infty}^{\infty}\;dz\:dv\;{\cal P}_E(z,v)
\;\delta\left(\tau+\frac{2v}{1+z^2}\right).
\end{equation}
The function ${\cal P}_E(z,v)$ in its turn can be extracted from
the correlation function
\begin{equation}\label{cr-fun}
{\cal K}_{E,\Omega}(z_1,z_2)=\left<\rho_{E-\Omega/2}(z_1)
\rho_{E+\Omega/2}(z_2)\right>-\left<\rho_{E-\Omega/2}(z_1)\right>
\left<\rho_{E+\Omega/2}(z_2)\right>
\end{equation}
of the densities $\rho_{E}(z)=(1/M)\sum_{a=1}^{M}\delta(z-z_a(E))$
of eigenvalues $z_a(E)$ at a given energy $E$ as:
\begin{eqnarray}
{\cal P}_E(z,v)&=&M\lim_{\Omega\to 0}\Omega\: {\cal K}_{E,\Omega}
(z_1=z-v\Omega/2,z_2=z+v\Omega/2)
\end{eqnarray}
 Finally, to get access to the
correlation function (\ref{cr-fun}) we introduce the related
object in terms of the traces of the resolvent:
\begin{equation}
\label{f}
f(z_1,z_2)=\left<{\rm Tr}\frac{1}{z_1-i\epsilon-\hat{K}(E-\Omega/2)}
\;{\rm Tr}\frac{1}{z_1+i\epsilon-\hat{K}(E+\Omega/2)}\right>,
\end{equation}
from which the required correlation function can be extracted as
\begin{equation}
{\cal K}_{E,\Omega}(z_1,z_2)=\frac{1}{2\pi^2 M^2}{\rm Re}\:f_c(z_1,z_2),
\;\;\;\; f_c(z_1,z_2)=f(z_1,z_2)-f^-(z_1)f^+(z_2).
\end{equation}

In order to calculate (\ref{f}) we obviously need to specify the
explicit form of the $K$-matrix. In the framework of the present
approach $\hat{K}$ can be written in terms of the random
Hamiltonian of a closed system represented by $L\times L$
Hermitian matrix $\hat{H}$, and some deterministic $L\times M$
matrix $\hat{W}$ describing the coupling of the system to the
leads, see \cite{VWZ85,Sokolov1989,Fyodorov1997}:
\begin{equation}
\label{K-matr} \hat{K}=\hat{W}^\dagger
\frac{1}{E-\hat{H}}\,\,\hat{W}.
\end{equation}
 An important
combination of these two matrices is the so-called effective
non-Hermitian Hamiltonian $\hat{H}_{eff}=\hat{H}-i\hat{\Gamma}$,
where $\hat{\Gamma}=\pi\hat{W}\hat{W}^\dagger$. Actually, for the
scattering problem $ \hat{H}_{eff}$ in many respects replaces the
Hamiltonian of the closed system $\hat{H}$.

There are several convenient microscopic models for the
Hamiltonian $\hat{H}$ describing the quasi-one-dimensional
disordered wire decoupled from leads. On may, for example, employ
the one-dimensional variant \cite{IWZ} of the gauge-invariant
N-orbital model introduced by Wegner \cite{SW}. Here we prefer to
work within the ensemble of $L\times L$ Banded Random Matrices
(BRM), see \cite{FM94Int,Haake} for detail and discussions. In the
limit of large number of orbitals $N\gg 1$ and of the large widths
of the band $b\gg 1$ both microscopic models can be reduced to the
same field-theoretical construction known as the nonlinear
$\sigma-$model and are therefore equivalent. For the sake of
simplicity in the main part of this section we consider explicitly
the case of Hermitian matrices corresponding to systems with
broken time reversal invariance due to a magnetic field. The
calculation for the case of preserved time reversal invariance is
similar in spirit, although more involved technically, and we
quote the corresponding results in the very end.

 The matrix
elements of the Hamiltonian $H_{ij}, i\le j$ are chosen to be
independent random variables with the joint probability density
\begin{equation}
\label{brm} {\mathcal P}(H) = {\mathcal
N}\prod_{i=1}^L\exp\left(-\frac{H_{ii}^2}
{2V_{ii}}\right)\prod_{i<j}\exp\left(-\frac{|H_{ij}|^2}{2V_{ij}}\right),
\end{equation}
where the variances $V_{ij}$ decay rapidly outside a band of width
$2b$ around the main diagonal:
\begin{equation}
V_{ij}=\frac{\lambda^2}{2b}\exp\left(-\frac{|i-j|}{b}\right)
\end{equation}
This ensemble was studied intensively during last years and many
analytical results are known in the limit $L\gg b \gg 1$ in the
closed form \cite{FM94Int}. As to the matrix
$\hat{\Gamma}\equiv\pi\hat{W}\hat{W}^\dagger$ describing coupling
to a single $M-$channel lead attached to the edge of the sample,
it can be verified that in the limit $L\gg b \gg M$ that matrix
can be chosen diagonal with only first $M$ elements different from
zero: $\hat{\Gamma}={\rm
diag}(\gamma_1,\dots,\gamma_M,0,\dots,0)$. The eigenvalues
$\gamma_a$ define the strength of coupling to a given channel in
the lead, and will enter the final expression via the coupling
coefficients $g_{\gamma_a}=2/T_a-1$, see Eq.(\ref{39}) below. The
perfect coupling corresponds as usual to $g_{\gamma_a}=T_a=1$.

To find the correlation function (\ref{f}) we express it via the
generating function:
\begin{eqnarray}
\label{gen-fun}
f(z_1,z_2)= \frac{\partial^2}{\partial J_1\partial J_2}\left.\left[\left(\frac
{Z_J^{(1)}Z_J^{(2)}}{Z_{J=0}^{(1)}Z_{J=0}^{(2)}}\right)^M {\cal F}(J_1, J_2)\right]
\right|_{J_1=J_2=0},\nonumber\\
 {\cal F}(J_1, J_2)=\left<\frac{{\rm Det}[E-\Omega/2-H_{eff}(Z_J^{(1)})]
\:{\rm Det}[E+\Omega/2-H_{eff}(Z_J^{(2)})]}{{\rm Det}[E-\Omega/2-H_{eff}
(Z_{J=0}^{(1)})]\:{\rm Det}[E+\Omega/2-H_{eff}(Z_{J=0}^{(2)})]}\right>,
\end{eqnarray}
where we introduced the notations: $Z_J^{(p)}=z_p+(-1)^p\epsilon
+J_p,\;\; p=1,2$ and defined  the effective Hermitian matrices
\begin{equation}
H_{eff}(Z_J^{(p)})=\hat{H}+\frac{1}{Z_j^{(p)}}\hat{\Gamma}
\end{equation}

The ensemble averaging in Eq.(\ref{gen-fun}) can be easily
performed after one represent the determinants in terms of the
Gaussian supersymmetric integrals\cite{Efbook}. We follow
notations in \cite{Fyodorov1997}, see also \cite{Haake}:
\begin{eqnarray}
\label{super-int}
{\cal F}(J_1, J_2)=(-1)^L\int \left[d\Psi\right]\:\exp\left\{-iE\Psi^{\dagger}
\hat{L}\Psi - i \frac{\Omega}{2}\Psi^{\dagger}\hat{L}\hat{\Lambda}\Psi+
 i\Psi^{\dagger}\hat{\Gamma}\otimes(\hat{L}\hat{U})\Psi\right\}
\left<\exp\left\{i\Psi^{\dagger}(\hat{H}\otimes\hat{L})\Psi\right\}\right>,
\end{eqnarray}
where:
\begin{eqnarray}
\Psi=\left(
\begin{array}{l}
\vec{\Psi}_{1} \\
\vec{\Psi}_{2}
\end{array}
\right), \;\;\;\; \mbox{where}\;\;\;\;
\vec{\Psi}=\left(
\begin{array}{l}
\vec{S}_{p} \\
\vec{\chi}_{p}
\end{array}
\right), \;\;\;\;p=1,2
\end{eqnarray}
The elements of the $L$-dimensional vectors $\vec{S}_{p}$ and $\vec{\chi}_{p}$
are commuting and anticommuting variables respectively. The $4\times 4$
matrices appearing in Eq.(\ref{super-int}) are diagonal $\hat{L}={\rm diag}
(1,1,-1,1),\;\hat{\Lambda}={\rm diag}(1,1,-1,-1),\;\hat{U}^{-1}={\rm diag}
(Z_{J=0}^{(1)},Z_{J}^{(1)},Z_{J=0}^{(2)},Z_{J}^{(2)})$.

In order to calculate right-hand side of Eq.(\ref{super-int}) we
perform all the standard steps \cite{FM94Int,Haake}: (i) average
over the random matrices $\hat{H}$ according the distribution
function (\ref{brm}); (ii) introduce auxiliary supermatrices
$\hat{Q}_i$  that allow to decouple the obtained integral by the
Hubbard-Stratonovich transformation; (iii) perform Gaussian
integral over $\Psi$; (iv) employ the saddle-point approximation
for the integrals over $Q_i$ justified by large $b$ and $L$. All
these steps allow us to represent the main object of interest as:
\begin{eqnarray}
\label{sigma}
f(z_1,z_2)=\lim_{J_{1,2}\to 0}\frac{\partial^2}{\partial J_1 \partial J_2}\int
\prod_{i=1}^{L}d\mu(\hat{Q}_i)\;\exp\left[-\frac{\xi}{4}\sum_{i=1}^L{\rm
Str} \hat{Q}_i\hat{Q}_{i+1} - i\frac{\Omega}{2}\pi\nu\sum_{i=1}^L{\rm Str}
\hat{Q}_i \hat{\Lambda}\right]\prod_{a=1}^M f_a,\\
f_a={\rm
Sdet}^{-1}\left[U^{-1}-\frac{\gamma_a}{V_0}\left(\frac{E}{2}-i\pi\nu
\hat{Q_a}\right)\right],\;\;\;\; V_0=\sum_{i=1}^LV_{ij}.
\end{eqnarray}
Here the supermatrices $\hat{Q}_i$ satisfy the following
constraints:
\begin{equation}
\hat{Q}_i^2=\hat{I},\;\;\;\;\hat{Q}_i^{\dagger}=\hat{L}\hat{Q}_i\hat{L}
,\;\;\;\; {\rm Str}\:\hat{Q}_i=0.
\end{equation}
The mean density of states $\nu$ around a point $E$ in the
spectrum is given in this limit by the semicircular law:
$\nu=\frac{1}{2\pi V_0}\sqrt{4V_0-E^2}$, whereas the parameter
$\xi = ((\pi\nu V_0)^2 8b\lambda^{-2}e^{-1/b})/(1-e^{-2/b}) \sim
b^2$ is nothing else but the localization length of the wire
detached from the leads \cite{FM94Int}. The integral over
supermatrices $Q_i$ can be calculated using the transfer matrix
technique \cite{FM94Int,Haake}. The technical details of the
calculation are rather similar to those performed in Refs
\cite{FM94Int,Fyodorov1997}, and we skip them here in favor of
presenting the final result for the density of the (scaled)
partial time delays $\ptime=\tau/2\pi\nu$:
\begin{equation}
\label{distr} {\cal
P}_{\tilde{\tau}}(\tilde{\tau})=\frac{(-1)^{M+1}}{2\pi
M!\:L^{M+1}\:
\ptime^{M+2}}\int_0^{2\pi}d\theta\;\left(g_{\gamma}-\sqrt{g_{\gamma}^2-1}\cos\theta\right)^M
\:w_L^{(M+1)}\left(\frac{g_{\gamma}-\sqrt{g_{\gamma}^2-1}\cos\theta}{\ptime
L}\right),
\end{equation}
Here the notation $w_L^{(M+1)}(z)\equiv
\frac{d^{M+1}w_L(z)}{dz^{M+1}}$ stands for $(M+1)$-th derivative
of the function $w_L(z)$ with respect to its argument.
 The function $w_L(z)$ is expressed in terms of the solution of the following
differential equation:
\begin{eqnarray}
\label{diff-ur} \frac{\partial W}{\partial
t}(y,t)&=&\left(y^2\frac{\partial^2}{\partial y^2}-y \right)
W(y,t),\;\;\;\;\;\; W(y,0)=1
\end{eqnarray}
as $w_L(z)=W(\xi z,L/\xi),$. In deriving the expression
Eq.(\ref{distr}) we assumed for simplicity  that all the
scattering channels have equal coupling strength $\gamma_a=\gamma$
related to the parameter $g_{\gamma}$ as:
\begin{equation}\label{39}
g_{\gamma}=\frac{1}{2\pi\nu\gamma}\left(\frac{\gamma^2}{V_0}+1\right).
\end{equation}

For the important case of perfect coupling to leads $g_{\gamma}=1$
, see the discussion around Eq.(\ref{coupling}), and the
expression (\ref{distr}) can be considerably simplified:
\begin{equation}
\label{perf}
 {\cal P}_{\tilde{\tau}}(\tilde{\tau})=\frac{(-1)^{M+1}}{ M!\:L^{M+1}\:
\ptime^{M+2}}\:w_L^{(M+1)}\left(\frac{1}{\ptime L}\right).
\end{equation}
According to the results of Ref.\cite{MF93} the distribution of
the eigenfunction intensities $y=V|\psi_n({\bf r})|^{2}$  at the
edge of the sample can be expressed in terms of the same function
$w_L(z)$ as:
\begin{equation}
\label{inten} {\cal
P}_y(y)=\frac{1}{L^2}w_L^{(2)}\left(\frac{y}{L}\right)
\end{equation}
Comparing these two expressions we conclude that
\begin{equation}\label{M}
 {\cal P}_{\tilde{\tau}}(\tilde{\tau})=\frac{(-1)^{M+1}}{ M!\:
\ptime^{M+2}}\:{\cal P}_y^{(M-1)}\left(\frac{1}{\ptime }\right)
\end{equation}
In particular for $M=1$ we rediscover Eq.(\ref{9}) verifying this
relation for a quasi-one-dimensional geometry in an independent
way. The exact solution of Eq.(\ref{diff-ur}) can be written in
terms of the modified Bessel functions \cite{FM94Int} and for a
single-channel setup it reproduces the result obtained by one of
the authors in a different way \cite{F03}.

Let us briefly discuss some asymptotic behaviour of the found
solution in a few important limiting cases. For the case of a
short wire, i.~e. $L/\xi \ll 1$ the solution of Eq.(\ref{diff-ur})
can be approximated by $W(y,t)\approx e^{-yt}$ so that
$w_L(z)=e^{-Lz}$. This approximation yields the probability
density for delay times
\begin{equation}
 {\cal P}_{\tilde{\tau}}(\tilde{\tau})=\frac{1}{ M!\:
\ptime^{M+2}}\:e^{-1/\ptime},
\end{equation}
which coincides with the distribution for  ``zero-dimensional''
case derived in the framework of the standard RMT
\cite{Fyodorov1997b,Gopar1996,Fyodorov1997}. Expanding the exact
solution in powers of the small parameter $1/g=L/\xi$ proportional
to the inverse dimensionless conductance for quasi one-dimensional
sample. We find:

\begin{equation}
 {\cal P}_{\tilde{\tau}}(\tilde{\tau})=\frac{1}{ M!\:
\ptime^{M+2}}\:e^{-1/\ptime}\left(1+\frac{1}{3g}\left[M(M+1)-2(M+1)\frac{1}
{\ptime}+\frac{1}{\ptime^2}\right]+\dots\right).
\end{equation}
For $M=1$ this formula gives back the expression
Eq.(\ref{metallic}), after we use that the value of the diffusion
propagator $\kappa$ is exactly equal to $2/3g$ at the edge of the
quasi-one-dimensional sample. The relation holds for not too small
values of delay times $\ptime>L/\xi$. In the opposite limit the
distribution is dominated by anomalously localized
states\cite{M00} and has the following form:
\begin{equation}
\label{asymp}
 {\cal P}_{\tilde{\tau}}(\tilde{\tau})\sim\frac{1}{\ptime^{(M+3)/2}}
\exp\left(-4\sqrt{\frac{\xi}{L\ptime}}\right)
\end{equation}
In the case of a long enough wire ($L>>\xi$) the probability
density  function is determined by the stationary
($t$-independent) solution of Eq.(\ref{diff-ur}) and is given
by
\begin{equation}
{\cal P}_{\tilde{\tau}}(\tilde{\tau})\approx\frac{(-1)^{M+1}}{
M!\: \ptime^{M+2}}\:\frac{8\xi^2}{L^2}
\frac{\partial^{(M-1)}}{\partial
\left[2\sqrt{\xi/{(L\ptime)}}\right]^{(M-1)}}
\left[K_1^2\left(2\sqrt{\xi/{(L\ptime)}}\right)+
K_0^2\left(2\sqrt{\xi/{(L\ptime)}}\right)\right].
\end{equation}
It has the same asymptotic behavior as Eq.(\ref{asymp}).

Finally let us briefly discuss the case  of the Hamiltonian matrix
$\hat{H}$ real and symmetric, which is pertinent for the systems
respecting the time-reversal invariance. It turns out that
 the above derivation can be straightforwardly generalized to this
 case. We omit all the cumbersome details of intermediate
 calculation and proceed directly to the final
 result for the density of partial time delays for $M$ equivalent
 channels and zero energy $E=0$:
\begin{equation}
{\cal
P}_{\tilde{\tau}}(\tilde{\tau})=\frac{C_M}{\ptime^2}\int_{-\infty}^{\infty}
dz\:\frac{1}{1+z^2}\int_{-\infty-i0}^{\infty-i0}\:dx\frac{w_L^{(2)}(ix)}{\left[
1-\frac{2i\gamma L (z^2+1)\ptime x}{z^2+\gamma^2}\right]^{M/2}},
\end{equation}
Here we use the same notation as in Eq.~(\ref{distr}) and $C_M$ is
a normalization constant, with the infinitesimal shift $-i0$
ensuring the integral to be well-defined. The expression can be
again simplified for the perfect coupling ($\gamma=1$):
\begin{equation}
\label{goe-perf} {\cal P}_{\tilde{\tau}}(\tilde{\tau})=\frac{\pi
C_M}{\ptime^2}\int_{-\infty-i0}^
{\infty-i0}dx\:\frac{w_L^{(2)}(ix)}{\left[ 1-2i L \ptime
x\right]^{M/2}},
\end{equation}
We shall consider separately the case of even and odd channel
number $M$. For $M=2p$, with $p$ being an integer we rewrite the
above expression as:
\begin{equation}
\label{goe-perf1} {\cal P}_{\tilde{\tau}}(\tilde{\tau})=\frac{\pi
C_p}{\ptime^{2+p}}\frac{(-1)^{p-1}}{(p-1)!(2L)^p}\frac{d^p}{dz^p}F(z)|_{z=\frac{1}
{2L\tilde{\tau}}} , \quad F(z)= \int_{-\infty-i0}^
{\infty-i0}dx\:\frac{w_L^{(2)}(ix)}{ z-ix}
\end{equation}
One can show that the function $w_L^{(2)}(ix)$ is analytic in the
lower half plane of the complex variable $ix$, and therefore the
integral in (\ref{goe-perf1}) is given by the residue at the pole
$x=-\frac{i} {2L\tilde{\tau}}$. Using the properties $w_L(0)=1$
and $w_L(\infty)=0$ to fix the normalization constant, we finally
arrive at the following simple expression:
\begin{equation}
\label{goe-perf2} {\cal
P}_{\tilde{\tau}}(\tilde{\tau})=\frac{1}{\ptime^{2+p}}\frac{(-1)^{p-1}}{p!(2L)^{p+1}}
\frac{d^{p+1}}{dz^{p+1}}w_L(z)|_{z=\frac{1} {2L\tilde{\tau}}}
\end{equation}
generalizing Eq.~(\ref{perf}) to the case of preserved time-reversal
invariance, and even number of channels.

For the odd number of channels $M=2p+1$ we use a very similar
manipulation and rewrite (\ref{goe-perf}) as:
\begin{equation}
\label{goe-perf3} {\cal P}_{\tilde{\tau}}(\tilde{\tau})=\frac{\pi
C_p}{(2L)^{\frac{M}{2}}\ptime^{2+\frac{M}{2}}}\frac{(-2)^{p}}{(2p-1)!!}
\frac{d^p}{dz^p}F(z)|_{z=\frac{1} {2L\tilde{\tau}}} , \quad F(z)=
\int_{-\infty-i0}^ {\infty-i0}dx\:\frac{w_L^{(2)}(ix)}{\sqrt{
z-ix}}
\end{equation}
The integral featuring in this expression can be reduced to that
along the branch cut parameterized as $x=-iq-\frac{i}
{2L\tilde{\tau}}$, where $0\le q< \infty$. After restoring the
normalization constant we find:
\begin{equation}
\label{goe-perf4} {\cal P}_{\tilde{\tau}}(\tilde{\tau})=
\frac{(-1)^{p}}{2^{p+3/2}\sqrt{\pi}\Gamma(p+3/2)}
\frac{1}{L^{p-\frac{1}{2}}\ptime^{p+\frac{5}{2}}}
\frac{d^p}{dz^p}F(z)|_{z=\frac{1} {2L\tilde{\tau}}} , \quad F(z)=
\int_{0}^ {\infty}\frac{dq}{\sqrt{q}}w_L^{(2)}(q+z)
\end{equation}
Since the one-channel case corresponds to $p=0$ we should expect
that the corresponding distribution of time delays is related by
Eq.(\ref{9}) to known distribution of eigenfunction intensities
found in \cite{MF93}, see also Eq.(3.14) of \cite{M00}. It is easy
to see that this is indeed the case. For arbitrary $p>0$ the time
delay density  can be also related to eigenfunction statistics by
differentiation, similarly to Eq.(\ref{M}). Finally, for the case
of a short wire we can use again  the approximate solution of the
differential equation (\ref{diff-ur})
 $W(y,t)\approx e^{-yt}$, so that $w_L(z)=e^{-Lz}$ and
 the integral in Eq.(\ref{goe-perf}) can
be calculated straightforwardly for all $M$, both odd and even.
This yields the density of partial time delays well-known from the
standard RMT \cite{FSS97}:
\begin{equation}
{\cal P}_{\tilde{\tau}}(\tilde{\ptime})=\frac{1}{2^{(M/2+1)}\Gamma(M/2+1)
\tau^{M/2+2}}\exp\left[-\frac{1}{2\ptime}\right]
\end{equation}

\section{Discussion and Conclusions.}
The main message of this paper is that measuring scattering
characteristics such as time delays one can get a direct access to
properties of eigenfunctions of the closed counterpart of the
scattering system.  For the simplest case of the single-channel
reflection from a disordered/chaotic sample the relation is most
explicit and is summarized in Eq.(\ref{9}), see also
Eq.(\ref{coupling}) below. It is very general (see discussion
below), and is valid for all pure symmetry classes (and even in
the crossover regimes, see Eqs.(\ref{cross})) and in the whole
parameter region ranging from fully localized to delocalized
eigenstates via the critical region of the Anderson transition.

In fact, one could suspect the existence of some relation of this
sort already from the original derivation\cite{Gopar1996} of time
delays distribution for the simplest case of zero spatial
dimension (RMT regime), see Eq.(\ref{rmt}). Indeed, the derivation
was based on relating the time delay statistics to that of
residues of the K-matrix by employing the so-called Wigner
conjecture. According to RMT those residues are chi-squared
distributed as a consequence of eigenfunction fluctuations
Eq.(\ref{chi2}). However, the authors \cite{Gopar1996} failed to
trace the generality, broad validity and physical consequences of
the ensuing relation beyond that simplest case.

The problem of relating scattering characteristics to the
properties of closed systems  also enjoyed some discussion in the
framework of a discrete-time evolution of the so-called network
models for disordered electron transport, see e.g. \cite{KZ} for
discussion and further references. Those models do not use a
Hamiltonian as the starting point, but rather operate with
networks of unidirectional links serving originally to mimic
electron propagation in strong magnetic field of Quantum Hall
phenomenon. After the present paper had already been submitted for
publication, we were informed by J.T.Chalker that a relation
analogous to our Eq.(\ref{9}) between the current density in a
link and the energy derivative of the total phaseshift emerged in
a one-dimensional version of the network model considered in
\cite{Chalker}.

Let us briefly emphasize a few points related to the validity of
the central formula Eq.(\ref{9}) which may need clarification.
First of all, statistics of the eigenfunction intensities entering
that relation refers to the position ${\bf r}$ of the lead/antenna
port, so that both quantities should be considered locally. This
may be important if the antenna attached not to the bulk, but
rather to the boundary of the sample, where eigenfunction
statistics may be modified due to boundary effects, see e.g.
\cite{M00}. Second, the derivation given in the sec. II of the
paper implies two very general assumptions only: i) The energy $E$
of the spectrum is not a singular point of the density of states,
so that the latter can be approximated by the
 constant value in the energy window of the order of the mean level
spacing. This condition is violated e.g close to the spectral
edges, or close to the band centre for systems with special (e.g.
"chiral", see \cite{FO04} and references therein) symmetries; and
ii) The phase of the scattering matrix is statistically
independent from the modulus of the scattering matrix and is
uniformly distributed over the unit circle\cite{FyodorovSavin04}.
The latter property is a very general feature of the {\it perfect
coupling} regime defined as one for which the energy/ensemble
 average $\langle S\rangle$ of the S-matrix vanishes.
 General case of non-perfect
 coupling is characterized by the so-called transmission
 coefficient $T=1-\mid<S>\mid^2<1$.
 \cite{VWZ85,Sokolov1989}, and it is well-known that
 statistics of various quantities at the perfect coupling $T=1$ allows
 one to find the properties for a general case $T\le 1$ after simple
 manipulations. In particular, such a relation for the partial delay
 times at arbitrary coupling $T$ can be found in\cite{SFS001}, see also\cite{Gopar1998}.
It allows one to write down the expression for the delay time
distribution for any coupling in terms of
 the statistics of the eigenfunction intensities as (cf. Eq.(\ref{distr}):
\begin{equation}\label{coupling}
\Pt=\frac{1}{2\pi}\int_0^{2\pi}
d\theta\:\frac{1}{2\pi\rho(\theta)\td^3} {\cal P}_y
\left(\frac{1}{2\pi\rho(\theta)\tilde{\tau}_w}\right),
\end{equation}
where the density of scattering phases
$\rho(\theta)=[2\pi(g_{\gamma}-\sqrt{
g_{\gamma}^2-1}\cos(\theta))]^{-1}$ is characterized by the
parameter $g_\gamma$\cite{foot2} related to the coupling strength
$g_{\gamma}\equiv 2/T-1$. For the relation of the parameter
$g_\gamma$ to "microscopic" characteristics of the banded random
matrix model see Eq.(\ref{39}) of the present paper. In
particular, the relation (\ref{coupling}) implies that moments of
the inverse time delay are always proportional to the moments of
the eigenfunction intensity $y=V|\psi_n({\bf r})|^{2}$  as:
\begin{equation}\label{mom}
\left\langle \td^{-k}\right\rangle=\left\langle
y^{k+1}\right\rangle \, \frac{1}{2\pi}\int_0^{2\pi}d\theta
\left[2\pi\rho(\theta))\right]^{k+1} .
\end{equation}
The range of validity of the relations Eqs.(\ref{coupling},
\ref{mom}) for a given microscopic model of a disordered system is
the same as the range of validity of the nonlinear $\sigma-$model
description of the latter, provided the singularities in spectrum
are avoided.

For the case of more than one channel statistics of {\it partial}
time delays can be also related to the statistics of
eigenfunctions as is clear e.g. from Eq.(\ref{M}). Here, however
we so far were unable to provide the generality and clarity
achieved for the single-channel derivation. We also leave as an
interesting open question in which way the {\it proper} time
delays whose statistics was in much detail investigated in
\cite{B97} could be related to statistics of eigenfunctions. The
same question remains for the Wigner time delay for more than one
open channel.

Referring to critical regime and corresponding multifractality as
reflected in the statistics of time delays, let us mention that
recently the scaling of the second negative moment of the Wigner
delay time was studied numerically at the metal-insulator
transition \cite{MK04}. It was indeed found that the scaling is
anomalous, but in contrast to the behaviour $\langle \tau_w^{-2}
\rangle \sim L^{-2D_3}$ expected from (\ref{scale}) the authors
reported $\langle \tau_w^{-2} \rangle \sim L^{-D_2}$.  Precise
reasons for such discrepancy are not clear for us at the moment
and deserve a separate investigation. We would like only to
mention that (i) for the case of $3D$ Anderson model the Wigner
time delay was numerically calculated for a disordered sample
coupled to a very large number of channels $M\gg 1$. This limiting
case is not covered by the present work, and requires separate
consideration; (ii) it is an open question to which extent the
numerical value $D_2=1.7$ reported in \cite{MK04} can be used for
reliable comparison with eigenfunction properties. Indeed, it
differs considerably from the value $D_2=1.3\pm 0.05$ found in
\cite{MEM} after a careful data analysis on eigenfunction
statistics. Another point that deserves mentioning is that
analytical results for the anomalous scaling of the eigenfunctions
were obtained in earlier publications after performing the spatial
averaging over the whole sample. Thus one expects validity of Eqs.
(\ref{weak}), (\ref{pl}) after averaging the left-hand side over
all possible positions of the lead/antenna. Alternatively, one
must be
 sure that the point where the lead is attached is a
 ``representative'' one, meaning
that the local statistics at this point is the same as the global
one. The last condition is usually satisfied due to eigenfunction
ergodicity, but some special situations like a lead attached to
the boundary of the sample may require special care.

\section{Acknowledgements}
We would like to thank Tsampikos Kottos, Vladimir Kravtsov and
John  Chalker for useful discussions, and Alexander Mirlin and
Dmitry Savin for the critical reading of the manuscript. The work
at Brunel was supported by EPSRC grant EP/C515056/1 "Random
matrices and Polynomials: a tool to understand complexity".


\begin{thebibliography}{99}
\bibitem{footnote}There is a misprint in the formula Eq.(5) of
\cite{FSS97}, the correct factor in that formula should be
$\lambda_2^{(M+3)/2}$ rather than $ \lambda_2^{(M+3/2)}$.
\end{thebibliography}
\end{document}